\documentclass[final,3p,times,twocolumn]{elsarticle}
\usepackage{amssymb,amsmath,lineno}
\usepackage{pst-node,pstricks,graphicx}
\usepackage{booktabs}
\usepackage{xspace}

\usepackage{booktabs}
\usepackage{nicefrac}
\usepackage[color]{showkeys}
\usepackage{fancyhdr}

\usepackage{units}
\begin{document}                
\begin{frontmatter}
  \title{Detection of tau neutrinos  by  Imaging Air  Cherenkov Telescopes }

\author[a]{D.~G\'ora}

\author[a,b]{E.~Bernardini}


\cortext[cor1]{E-mail address of corresponding author: Dariusz.Gora@desy.de}

\address[a]{Institut f\"ur Physik, Humboldt-Universit\"at zu Berlin, D-12489 Berlin, Germany}

\address[b]{DESY, Platanenallee 6, D-15738 Zeuthen, Germany}

\begin{abstract}
This paper investigates the potential to detect tau neutrinos in the energy range of 1-1000 PeV searching for very inclined showers with imaging Cherenkov telescopes. A neutrino induced tau lepton escaping from the Earth may decay and initiate an air shower which can be detected by a fluorescence or Cherenkov telescope.  We present here a study of the detection potential of Earth-skimming neutrinos taking into account neutrino interactions in the Earth crust,
local matter distributions at  various detector sites, the development of tau-induced showers in air and the detection of Cherenkov photons with IACTs. We analyzed simulated shower images on the camera  focal plane and implemented generic reconstruction chains based on Hillas parameters. We find that  present IACTs can  distinguish  air showers induced by  tau neutrinos  from  the  background of hadronic showers in the PeV-EeV energy range. We present  the neutrino 
trigger efficiency obtained for a few configurations being considered for the next-generation Cherenkov telescopes,  i.e. the Cherenkov Telescope Array.  Finally, for a few representative neutrino  spectra expected  from astrophysical sources, we compare the expected event  rates at running  IACTs to what is expected for the dedicated IceCube neutrino telescope.

\end{abstract}
\end{frontmatter}

\section{Introduction}

The discovery of an astrophysical flux of high-energy neutrinos by IceCube~\cite{HESE2} is a major step 
forward in the ongoing search for the origin of cosmic rays, since the neutrino emission may be produced
by hadronic interactions in astrophysical accelerators.  Of particular interest is the identification
of $\nu_{\tau}$, which are only expected to be produced in negligible amounts in astrophysical  accelerators, but should appear in 
the flux detected by IceCube due to neutrino flavor change. Up to now, there has been no clear identification of  $\nu_{\tau}$ at high
energies, so  the detection  of $\nu_{\tau}$ neutrinos will be very important from astrophysical and the particle physics point of view. 
The detection would give new information about the astrophysical flux as well as serving 
as an additional confirmation of the astrophysical origin of the IceCube high energy diffuse neutrino signal. It also 
would shed light on the emission mechanisms at the source, test the fundamental properties of neutrinos over extremely long baselines and
better constrain new physics models which predict significant deviations from equal fractions of all flavors.

The  existing Imaging Air Cherenkov Telescopes (IACTs) such as MAGIC~\cite{magic}, VERITAS~\cite{veritas} and H.E.S.S.~\cite{hess} could have the capability to detect PeV tau neutrinos by searching for very inclined showers~\cite{fargion}. In order to do that, the Cherenkov telescopes need to be pointed in the direction of the taus escaping from the Earth crust, i.e.\ at or a few degrees below the horizon. In \cite{upgoing_magic}, the effective area for up-going tau neutrino observations with the MAGIC telescopes was calculated analytically and found to be maximum  in the range from 100\,TeV to $\sim$~1\,EeV. However, the  sensitivity for diffuse neutrinos was found to be  very low because of the limited field of view (FOV) (the topographic conditions allow to point the telescopes only for a small window of about 1 degree width in zenith and azimuth to point the telescope downhill), the observation time and the low expected neutrino flux.

On the other hand, if flaring or disrupting point sources such as GRBs are being pointed to, one can  expect an observable number of events even from a single GRB if close by, as recently  shown  by    the All-sky Survey High Resolution Air-shower (Ashra) team \cite{Asaoka:2012em}. Also, for  IACT sites with different topographic conditions, the acceptance for up-going tau neutrinos is increased by the presence of mountains~\cite{gora:2015}, which serve as target for neutrino interaction
 leading to an enhancement in the flux of emerging tau leptons. A target mountain can also shield against cosmic rays and star light. Nights with high clouds often prevent the observation of $\gamma$-ray sources, but still allow  pointing the telescopes to the horizon. 
As an example for the MAGIC site there are about 100 hours per year  where high clouds are present~\cite{clouds}, therefore a large  amount of data  can be possibly accumulated. 
While the observation of tau neutrinos is not the primary goal of IACTs, a certain level of  complementarity can be expected when  switching from normal (i.e. $\gamma$-ray) observations mode to tau neutrinos (i.e. mostly horizontal) pointing.
 Next-generation Cherenkov telescopes, i.e.\ the Cherenkov Telescope Array (CTA)~\cite{cta}, can in addition exploit their much larger FOV (in extended observation mode) and  a higher effective area.
 \begin{figure*}[t]                                                                                                                
  \begin{center}                                                                                     
    \includegraphics[width=0.69 \columnwidth]{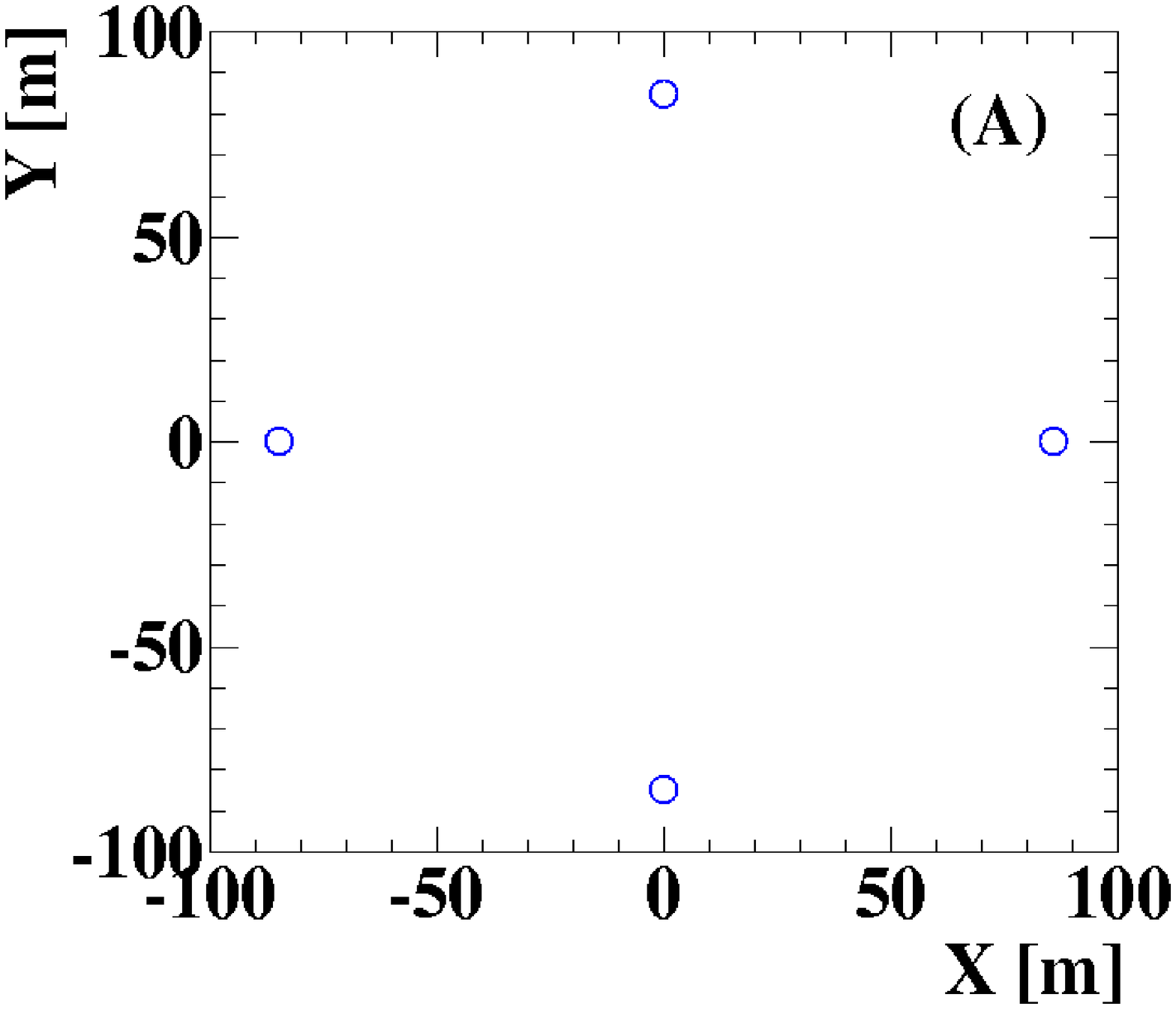}   
      \includegraphics[width=0.69 \columnwidth]{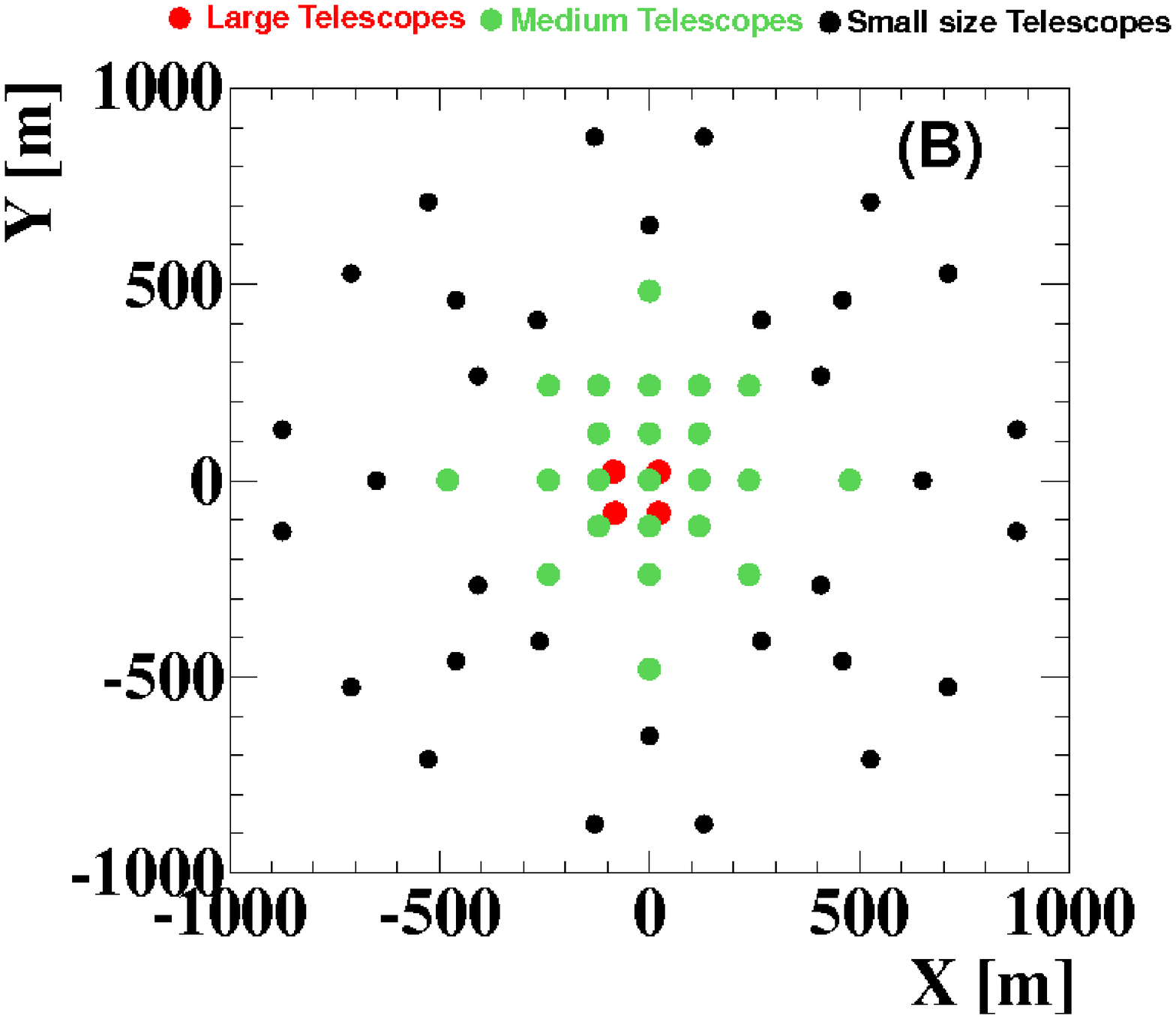}
       \includegraphics[width=0.69 \columnwidth]{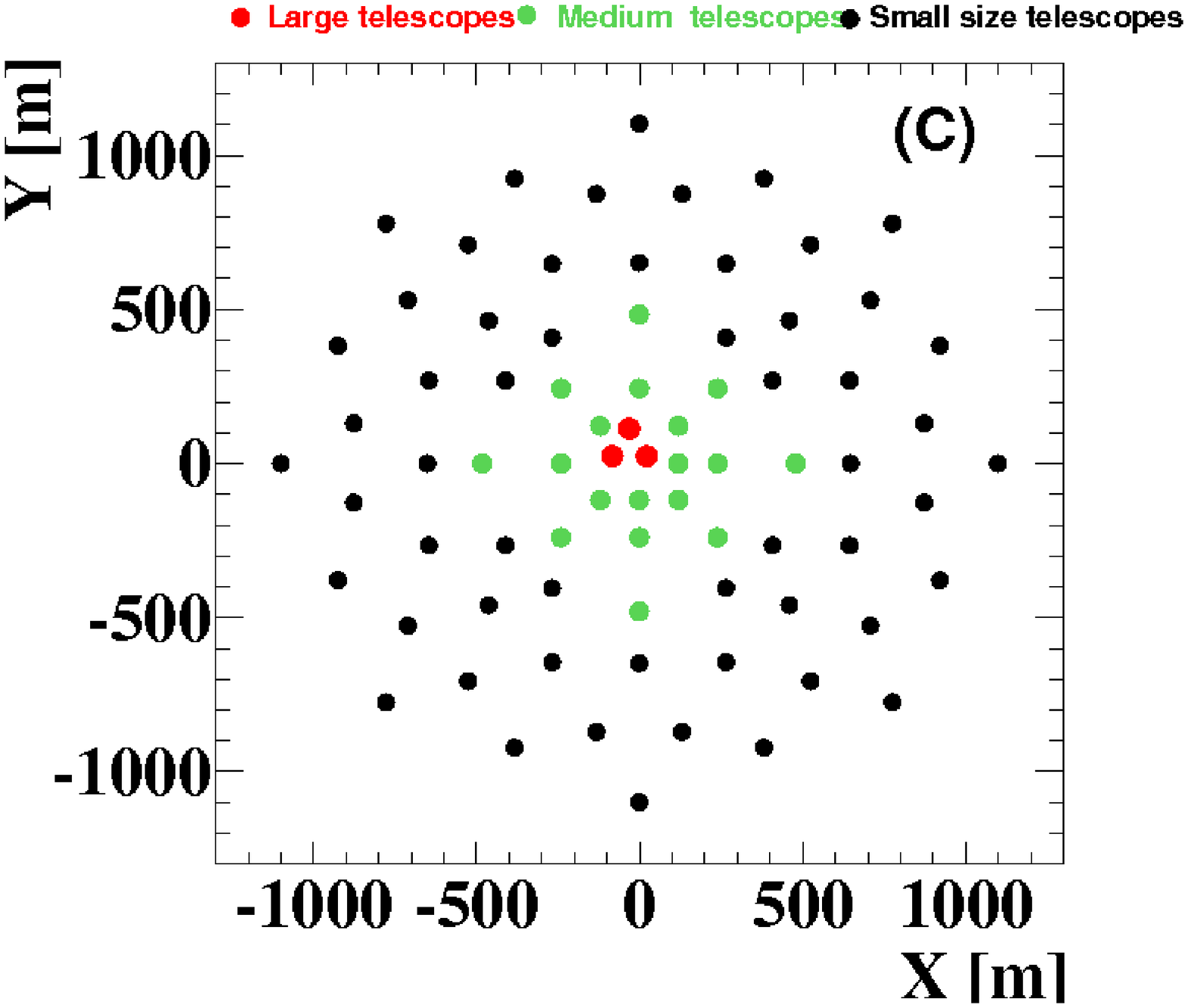}
  \end{center}
\caption{ \label{fig::cta::layout} {\bf Cherenkov telescope  layouts}  considered  in this work IACT-4 (A), CTA-E (B) and CTA-I (C). The IACT-4 array consists of
four Cherenkov Telescopes with   $\sim12$ m aperture, $2.5^{\circ}$ FOV and  $0.16^{\circ}$ camera pixel size, while  CTA arrays consist of  telescopes of different  size i.e. Large Size Telescopes (LST) with   $\sim23$ m aperture, $5^{\circ}$ FOV and  $0.09^{\circ}$ camera pixel size  (red full circles),  Medium Size Telescopes (MST)   with $\sim$ 12 m aperture,  $8^{\circ}$  FOV and $0.18^{\circ}$ camera pixel size  (open black circles) and  Small Size Telescopes  (SST) with $\sim$ 4-7 m aperture,  $10^{\circ}$  FOV and  $0.25^{\circ}$ camera pixel size (black full circles).   For  more detailed description of CTA telescope properties, see Table 1 in ~\cite{ctasim}.} 
\end{figure*}

In this paper,  we present an update of the work in~\cite{gora:2015},
where a detailed Monte Carlo (MC) simulation of event rates induced by Earth skimming tau neutrinos was performed
for an ideal Cherenkov detector in case of MAGIC, VERITAS site  and two proposed Cherenkov Telescope Array sites: Meteor Crater and Yavapai Ranch. 
For VERITAS and the considered Cherenkov Telescope Array sites the expected neutrino sensitivities are up to factor 3 higher than for the MAGIC site
because of the presence of surrounding mountains. The calculated neutrino rates are comparable to what has been estimated for the IceCube neutrino
telescope assuming realistic observation times for Cherenkov telescopes of a few hours.

However, in our previous work  the calculated event rate  were obtained with an assumed efficiency for tau induced shower of about 10\%. 
Here, we present a more detailed  simulation of trigger and identification efficiency for air showers induced by Earth-skimming tau neutrinos,
for  IACTs  and  for a  few  CTA  layouts considered in~\cite{ctasim}. We analyzed the simulated shower images on the camera focal plane showing that IACTs/CTA can distinguish air showers induced by tau neutrinos from the background of very inclined hadronic showers. We also recalculated the 
point source acceptance and the expected event rate taking into account this new estimation of the trigger efficiency. 

The structure of the paper is the following: Section~\ref{method}  describes our MC simulation chain. In Section~\ref{sec:results}  we show  the trigger/identification efficiencies for $\tau-$induced showers as a function of tau lepton energy and  we  study the  properties of  shower images  on the camera  focal plane,  as described by Hillas parameters.  This section  presents also an update of  our  previous work~\cite{gora:2015}.
Finally,  we  summarize  the results and give a conclusion in Section~\ref{summary}.

\section{Method}
\label{method}

In order to study the signatures expected from neutrino-induced showers by IACTs, a
full Monte Carlo (MC) simulation chain was set, which consists of three steps.

First, the propagation of a given neutrino flux through the Earth and the atmosphere is simulated using  an extended version of the ANIS code~\cite{gora:2007}. For fixed neutrino energies, $10^{6}$ events are generated on top of the atmosphere with zenith angles ($\theta$) in the range $90^{\circ}$--$105^{\circ}$ (up-going showers) and with azimuth angles in the range $0^{\circ}$--$360^{\circ}$. Neutrinos are propagated along their trajectories of length $ \Delta L$ from the generation point on top of the atmosphere to the interaction volume,  defined as the volume which can contribute to the expected event rate, in steps of $\Delta L$/1000 ($\Delta L/1000 \geq 6$ km). At each step of propagation, the $\nu$--nucleon interaction probability is calculated according to a parametrization of its cross section based on the chosen parton distribution function (PDF). In particular, the propagation of tau leptons through the Earth is simulated. All computations are done using digital elevation maps (DEM)~\cite{dem} to model the surrounding mass distribution of each site under consideration. The flux of the leptons emerging from the ground as well as their energy and the decay vertex positions are calculated inside an interaction volume, modeled by a cylinder with radius of 35\,km and height 10\,km, see also~\cite{gora:2007,gora:2015} for more details.

Then,   the shower development  of $\tau$-induced showers  and  Cherenkov light production  from such showers  is simulated with CORSIKA~\cite{corsika}.  CORSIKA   (version 6.99)   was compiled with the TAULEP option~\cite{taulep}, such that the tau decay is simulated with  PYTHIA~\cite{pythia}. In order  to simulate  Cherenkov light from  inclined showers for any defined Cherenkov  telescopes array the  CERENKOV and  IACT options were also activated~\cite{simtelarray}. Finally,  to consider  the atmospheric depth correctly for inclined showers, the  CURVED EARTH and SLANT options were also selected.
 Up to now, we could not simulate  showers with  zenith angle $\theta>90^{\circ}$ when combining  the "CURVED EARTH" and IACT options. Therefore, we use here a zenith angle of $87^{\circ}$ to estimate the trigger efficiency for up-going tau neutrino showers. This should be a reasonable assumption, because  the trigger efficiency  in case of  $\tau$-induced showers with the same energy  should only slightly depend on the zenith angle (as its confirmed by our later results),  as long as the corresponding altitudes of  shower maxima  are similar.

 The  CORSIKA  simulations were performed for different configurations: H.E.S.S. like   four telescopes 
(named here by  IACT-4), and  for  a few  CTA arrays  considered  in~\cite{ctasim}, see Figure~\ref{fig::cta::layout}. 
The IACT-4   can be considered as representative for current generation of IACTs.  Among  different CTA array configurations  shown in~\cite{ctasim}  the arrays  chosen were  named CTA-E  (59 telescopes) and CTE-I  (72 telescopes),  which according to ~\cite{ctasim} are the best compromise between compact   and  dense layout. The selected arrays  have only slightly worse  sensitivity for $\gamma$-rays  than the  full  CTA array~\cite{ctasim}.

We simulated showers induced by  tau leptons with energies from  1 - 1000 {PeV} in steps of 0.33 decades and   with an injection position at altitudes ranging from   detector  level to the  top of the atmosphere. We used  as the detector level 1800 a.s.l for the simulation of current generation  of  IACTs  and  2000 m a.s.l.  for  CTA. The injection point spans  different vertical depths  from  ground to top of the atmosphere with steps of at least  50 g/cm$^2$.  At each vertical depth,  1000 showers  were generated in order to  study shower-to-shower fluctuations and to cover  different tau decay channels. 
\begin{figure*}[t]                                                                                                             
\begin{center}                                                                                     
\includegraphics[width=0.8\columnwidth]{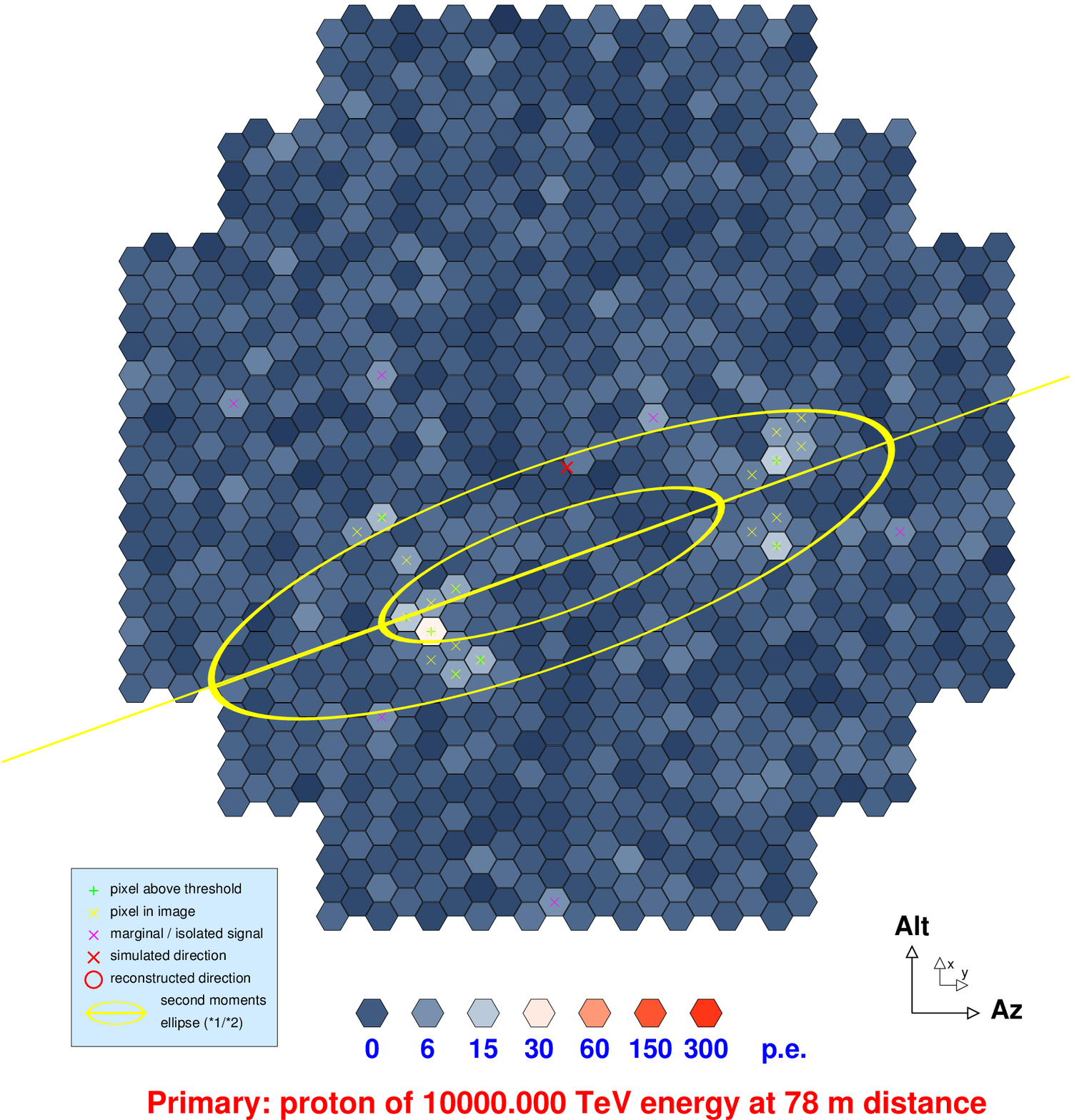}    
 \includegraphics[width=0.8\columnwidth,height=6.5cm]{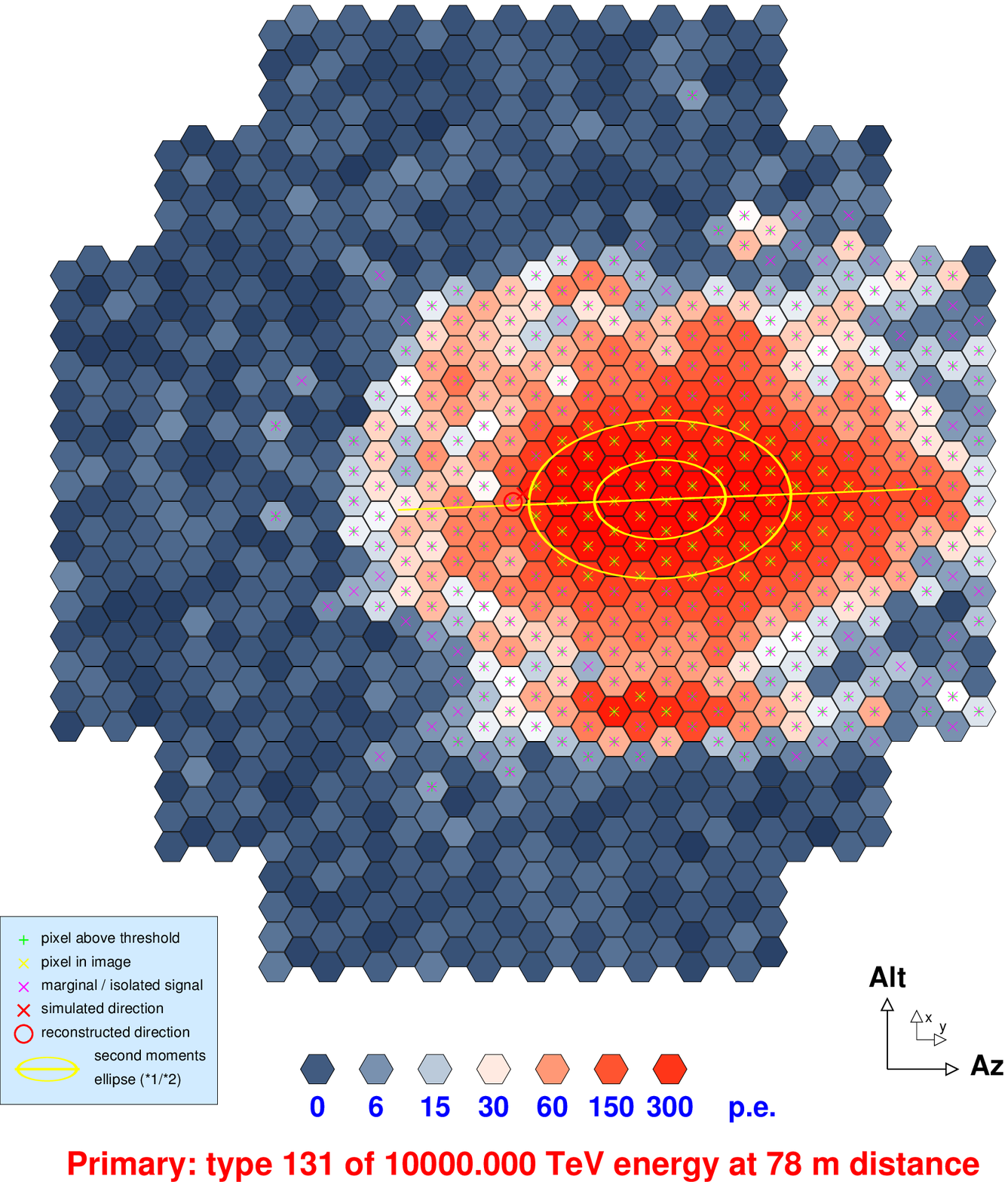}
\end{center}
\vspace{-0.5cm}
 \caption{\label{fig::hessimage} Example of simulated shower images with primary particle energy 10 PeV  and zenith angle $\theta=88^{\circ}$ as seen by a IACT-4 camera. (Left)  proton interacting at the top of the atmosphere,   first interaction point at vertical depth below 50 g/cm$^2$ and detector-to-shower distance of about  1000 km; (Right)  lepton  tau decaying close to the detector, with an  injection vertical depth of  760 g/cm$^2$ and a  detector-to-shower  distance  of about  50 km.} 
\end{figure*}   
For  each  CORSIKA  simulated shower  the impact point was randomized in a circle with radius $R_{max}$ on a plane perpendicular to the shower axis i.e. the CSCAT with VOLUMEDET/IACT option  was used. This radius was  optimized by looking to the fraction of triggered events as a function of $R_{max}$, and finally was set to $R_{max}=200$~m  for IACT-4 and $R_{max}=1000$~m  for CTA-E  in order to avoid information loss due to showers which could be triggered but were not simulated. 

For high energies ($>1$ PeV)  the computing time become excessively long (scaling roughly with the primary energy). In order to reduce it to tolerable values the so-called
"thin sampling" mechanism is used~\cite{thinning}. To cope with the vast number of secondary particles  thinning and re-weighting of secondaries was used with a thinning level of 10$^{-6}$ . The kinetic energy thresholds for explicit tracked particles were set to: 300, 100,  1,  1~MeV for hadrons, muons, electrons and photons, respectively. Shower simulations were performed considering  the QGSJET II  model  for hadronic interactions in the atmosphere.

The results of CORSIKA simulations were used as the input for the last step i.e. simulation of the detector response. We used  the Cherenkov telescope simulation package: {\tt sim\_telarray}~\cite{simtelarray}.  The light collection area is simulated  including the ray-tracing of the optical system, the measured transmittance and the quantum efficiency of PMTs. The response of the  camera  electronics  was simulated in detail  including   night-sky background and different system triggers.
 The {\tt sim\_telarray} simulations were performed for  IACT-4, and  for CTA-E and CTA-I with so-called {\it production-1} settings. 
The  response to   $\tau$-induced showers is found to depend weakly on the details of the optical set-up, field of view and  camera electronics.

In order to compare images at the camera plane we also simulated inclined showers induced by  protons, photons  and  electrons. At  energies larger than  1 PeV, we do not expect significant background of showers initiated by photons or electrons. The proton  simulations were instead  used to estimate the main isotropic background for  neutrino searches due to   interaction of comics rays  in the atmosphere. In order to have enough statistic  we use a similar strategy to the  case  of $\tau-$induced shower
i.e. we simulated proton induced showers with  primary  particle energy ranging from 1 to 1000 PeV in steps of 0.33 decades. At  each considered  zenith angle bins (80$^{\circ}$, 83$^{\circ}$, 85$^{\circ}$, 87$^{\circ}$)  the number of simulated events in CORSIKA input card was set  to the  corresponding number  of events  
 from the power law spectrum with   spectral index $\gamma=-2.7$. The direction of primary protons was varied   within a circle  with aperture   $\beta=5^{\circ}$ around the fixed primary direction, i.e. the VIEWCONE option was selected in the CORSIKA simulations.

\section{Results} \label{sec:results}
 
\subsection{Image on the camera}
In case of   showers observed at large zenith angles the Cherenkov light has to 
undergo a long optical path, due to  a thicker layer of the atmosphere. The shower maximum is located far
 from the observatory and the photon density  at the mirrors decreases. This reduces the efficiency compared to lower zenith angles, especially at low energies. Images on the camera  will be dimmer and smaller in size. 

As an example, in Figure~\ref{fig::hessimage}   we show  a representative  shower image for a 10~PeV proton  injected at the top of the atmosphere and a 10 
 PeV tau lepton injected  close to the detector, respectively.  As expected, the shower image on the focal camera plane for the tau lepton  has a much larger image  size  and contains much more photons  compared to the  proton one. Note also, that for inclined showers
the hadronic and electro-magnetic component is  almost completely  absorbed in the atmosphere while the muonic component (muons) can reach the Earth.
Thus, the showers images  on the cameras from $p$-induced showers  will  mostly contain the  muon ring (if muons propagate parallel to the optical axis) or   incomplete ring (arcs) in the camera, see Figure~\ref{fig::hessimage} (Left) as an example.
\begin{figure*}[t]                                                                                                          
  \begin{center}                                                                                     
    \includegraphics[width=\columnwidth,height=6.0cm]{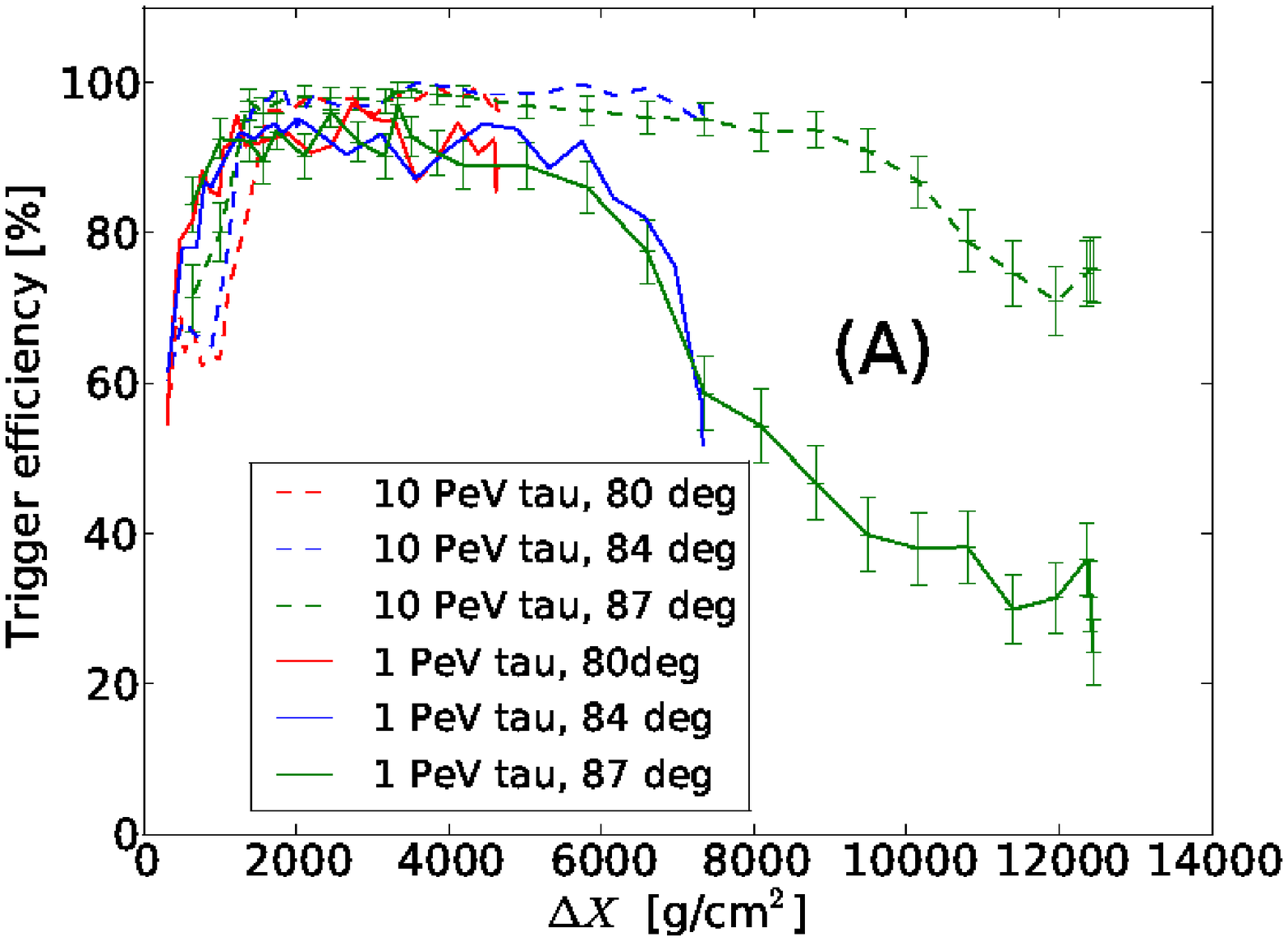}  
      \includegraphics[width=\columnwidth,height=6.0cm]{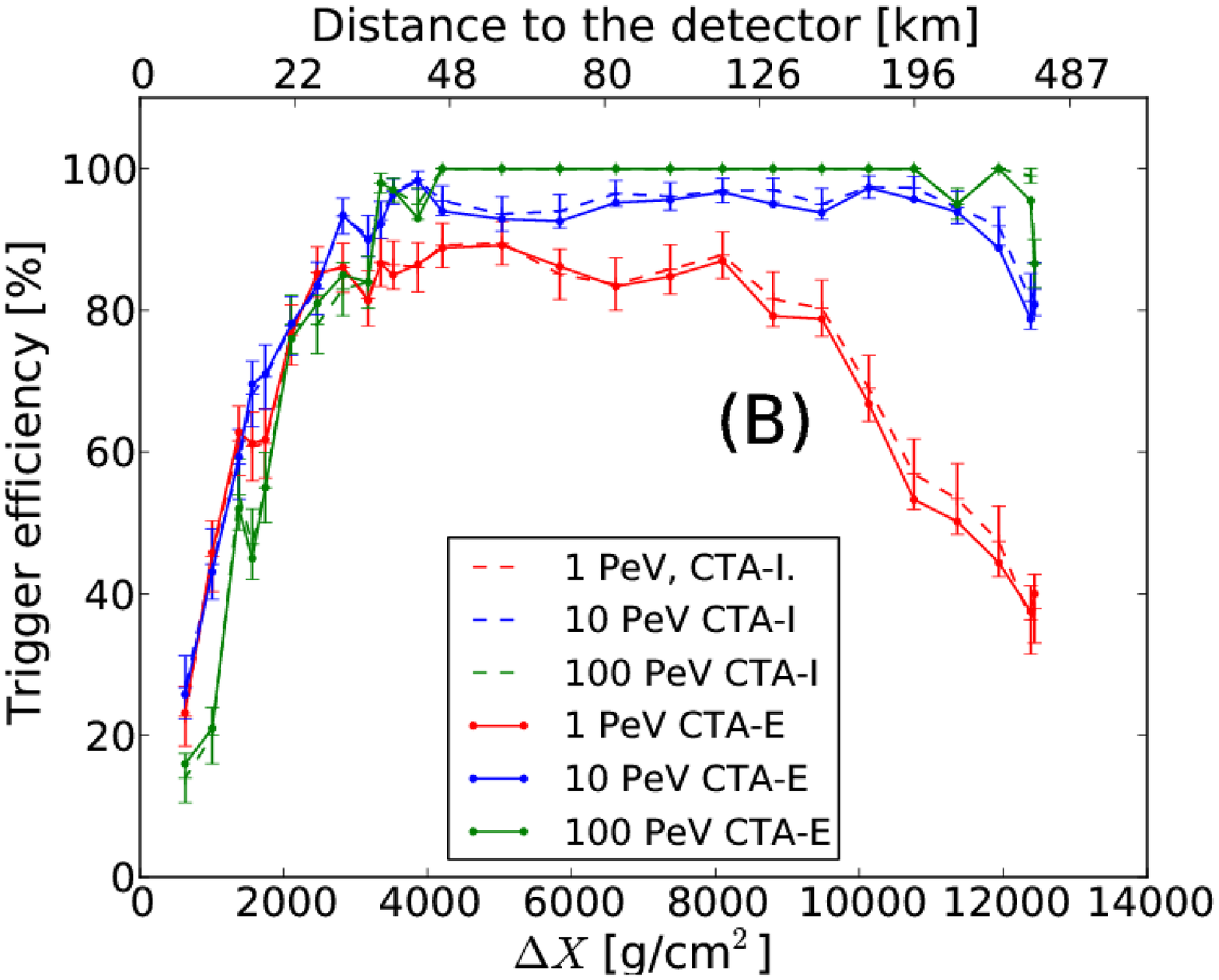}  
  \end{center}
 \caption{\label{trigger333}(A) Trigger efficiency as a function of the distance between injection point and the detector  level measured in g/cm$^2$ ($\Delta X$) with IACT-4  for different zenith angles and energies of the tau lepton.  Note, that for different zenith angles   the distance from the atmospheric border to detector level  is significant different due to the Earth's curvature.  (B) Trigger probability for  CTA at a fixed zenith angle of  $87^{\circ}$.The distance $\Delta X=0$ g/cm$^2$ corresponds to the detector level, $\Delta X\simeq12 000$ g/cm$^{2}$ to  the top of the atmosphere. }  
 \vspace{-0.5cm}
\end{figure*}

\subsection{Trigger efficiency}
The trigger efficiency (trigger probability) depends on the response of a given detector and
is usually estimated based on MC simulations. The trigger efficiency, $T(\theta, E_{i},X)$  in an energy range interval $\Delta E$, is defined 
as the  number of simulated showers with positive trigger decision over the total number of generated showers
for  fixed zenith angle $\theta$, initial energy of primary particle $E_{i}$ and injection depth~$X$.  In this work,   simulations were  done  for a two level  trigger, so-called Majority trigger. The first level  is a camera level  trigger ({\bf L1}) defined by  3 pixels  above 4 photo-electrons (p.e.)  within  a short time window and the second level is basically a coincidence level trigger among all telescopes in the  defined array or sub-array ({\bf L2}) and requires at least 2 neighboring  triggered telescopes.

Figure~\ref{trigger333} (A) shows  the trigger  probability ({\bf L2}) for $\tau$-induced showers with different zenith angles and  energies of the tau lepton in case of the IACT-4 array.  The  calculated trigger probabilities for different zenith angles $\theta=80^{\circ}, 84^{\circ}, 87^{\circ}$  are  quite similar, within errors,  if plotted as a function of the  distance between the injection point and the detector  measured in g/cm$^2$   (in this work this distance to the detector was labeled as  $\Delta X$). This is understood, if we note  that  amount of  Cherenkov light detected  depends essentially on the distance  between the Cherenkov telescope and  the shower maximum.  At its maximum  a shower has the largest lateral extension and  Cherenkov light production,  thus is capable of producing the largest signal  seen by IACTs  telescopes. 

 As expected (see Figure~\ref{trigger333} (A)) the trigger probability  increases with  primary energy of the tau lepton  and decreasing distance to the detector. Only,   at  $\Delta X < 1000$ g/cm$^2$,  the trigger efficiency drops due to the fact that the shower maximum is too close to the detector or the shower did not reach yet the maximum of shower development,  decreasing  the amount of Cherenkov light seen by telescopes. It is also worth to mention, that
below  $\Delta X < 6000$ g/cm$^{2}$     the trigger probability  is  at the level of about  90\%. In this case the  corresponding geometrical  distance to the detector (in meters) depends on  the zenith angle $\theta$, but  for $\theta=87^{\circ}$  is of  about   $\sim 80$ km. This  provides an estimate  of the size of the  active volume for $\tau$-induced showers seen by IACTs.  
\begin{figure*}[ht]                                                                                                                
  \begin{center}                                                                                      
     \includegraphics[width=2\columnwidth]{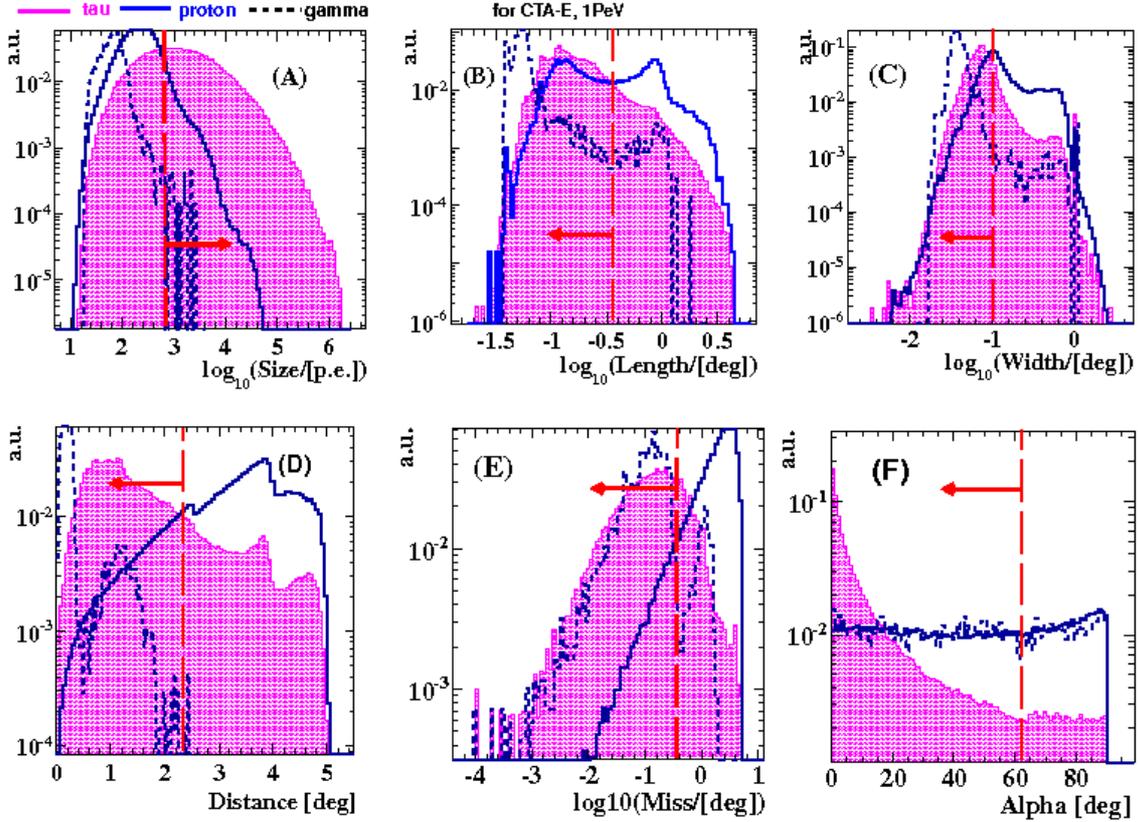} 
  \end{center}
 \caption{\label{fig::hillas} Normalized distribution of Hillas parameters  for $\tau$, $p$ and $\gamma$-induced showers,  zenith angle $\theta=87^{\circ}$ and CTA-E. Only  deep $\tau$-induced showers with $\Delta X <8800$ g/cm$^2$ and primary particle energy 1 PeV are shown, while for $p/\gamma$ only events  interacting at the top of the atmosphere with $\Delta X >11400$ g/cm$^2 $  are considered. The  $p$-events come from  CORSIKA simulations for  primary protons with energies between 1PeV and 1000 PeV with a differential spectral index of $-2.7$, while  $\gamma$-events from  simulations 
with the primary photon energy  of 1 PeV. Vertical dashed  lines and arrows indicate  our selection cuts developed for $\tau$-induced showers, see text for more details.} 
\end{figure*}   
 
  Figure~\ref{trigger333} (B) shows  the trigger probability for the considered  CTA arrays  shown in Figure~\ref{fig::cta::layout} (B) and (C)  and  different  primary  energy of lepton tau.  As  for IACT-4,  the trigger  probability increases because the higher is the energy, the more Cherenkov light is produced, and the larger the number of triggered events. Comparing
 with results from  Figure~\ref{trigger333}~(A) calculated for larger CTA arrays,  with  more telescopes with different optics and camera structures,
 we find basically a similar  fraction of triggered events (above $\Delta X>2000$ g/cm$^2$). The difference in the trigger probability  seen for  $\Delta X < 2000$ g/cm$^{2}$ between IACT-4 and CTA-E it  is due  to  the different  altitudes of detectors i.e.  a higher altitude for CTA-E. 
The altitude  difference   is only   200 m, but  for zenith angle $\theta=87^{\circ}$  it    translates into a difference of about 4 km  in the detector to shower distance. In case of IACT-4 this  leads to a  larger fraction of  triggered showers than for CTA-E, because   more showers  can reach their  maximum of shower development. 
 Moreover, for the considered  CTA arrays,  the trigger efficiency  only  slightly depend on the  array  structure.  This  can be explained by the fact, that for inclined showers  studied in this work (with $\theta>80^{\circ}$)    the  radius  of the Cherenkov light pool distribution  at detector level  is  larger than 1~km~\footnote{For index of refraction $n_{air}=1.00023$ at an altitude of 1800~m, the Cherenkov opening angle is $\alpha \simeq 1.2^{\circ}$.  Thus, for geometrical distance from the  shower maximum to detector
of  about   50 km the Cherenkov ring  radius on the ground,  assuming  not changes of refraction index within this distance, is given by:  50~km~$\times \tan(\alpha)/ \cos(\theta)$=1.04 km/{$\cos(\theta$)} km  for fixed zenith angle $\theta$.}, which  is much more  than  the distance between telescopes in the considered arrays. Thus, the fraction of triggered events is expected to be similar and to be only weakly dependent on the density of telescopes.
\begin{figure*}[ht]                                                                                                                
  \begin{center}                                                                                     
    \includegraphics[width=2.0\columnwidth]{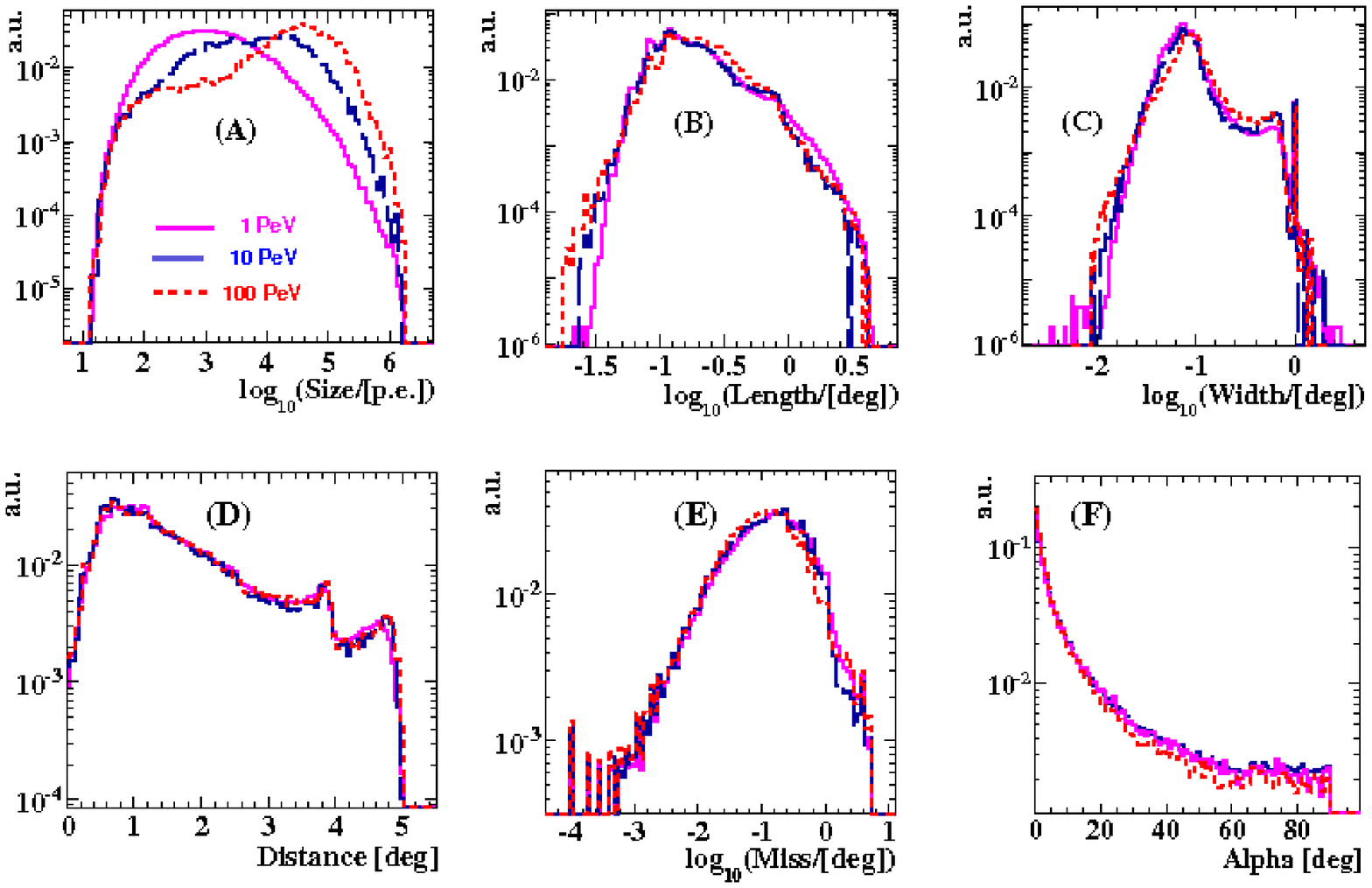}   
  \end{center}
 \caption{\label{fig::hillas2} Normalized distribution of Hillas parameters for 1, 10 and 100 PeV  in case of $\tau-$induced showers
 for CTA-E   array and   zenith angle $\theta=87^{\circ}$.  } 
\end{figure*}   
\subsection{Discrimination of $tau-$induced showers }

In this section  we show how to discriminate of $\tau-$induced showers from background hadronic  showers. The results presented here are based
on simulation of down-going showers with zenith angle $\theta>84^{\circ}$, but they  can be applied to any neutrino flavour,
since all neutrinos  with different flavours  can induced down-going air showers, which produce a large amount of Cherenkov light at high energies ($ > 1$ PeV).
We already show in Figure~\ref{trigger333} (A) that the trigger probability does not depend on zenith angle for inclined showers, 
thus  it  can be used  for down-going neutrino searches, as well. Of course,  in such a case  the neutrino sensitivity  is reduced due to small
target density for neutrino interaction (happening in the atmosphere),
compared to the sensitivity obtained for Earth-skimming neutrinos.

\begin{figure*}[t]                                                                                                             
  \begin{center}                                                                                  
    \includegraphics[width=\columnwidth,height=5.5cm ]{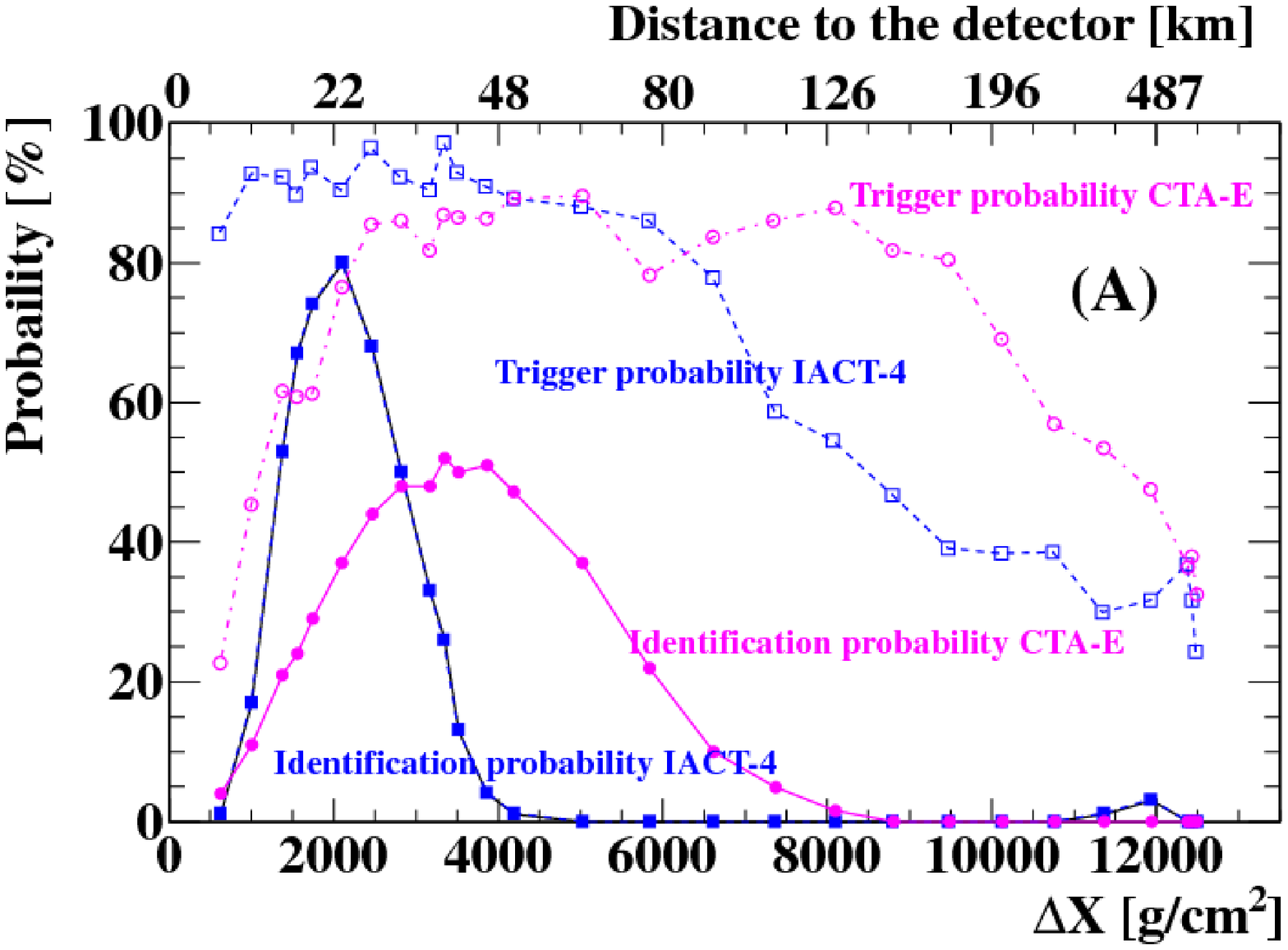}   
     \includegraphics[width=\columnwidth,height=5.5cm]{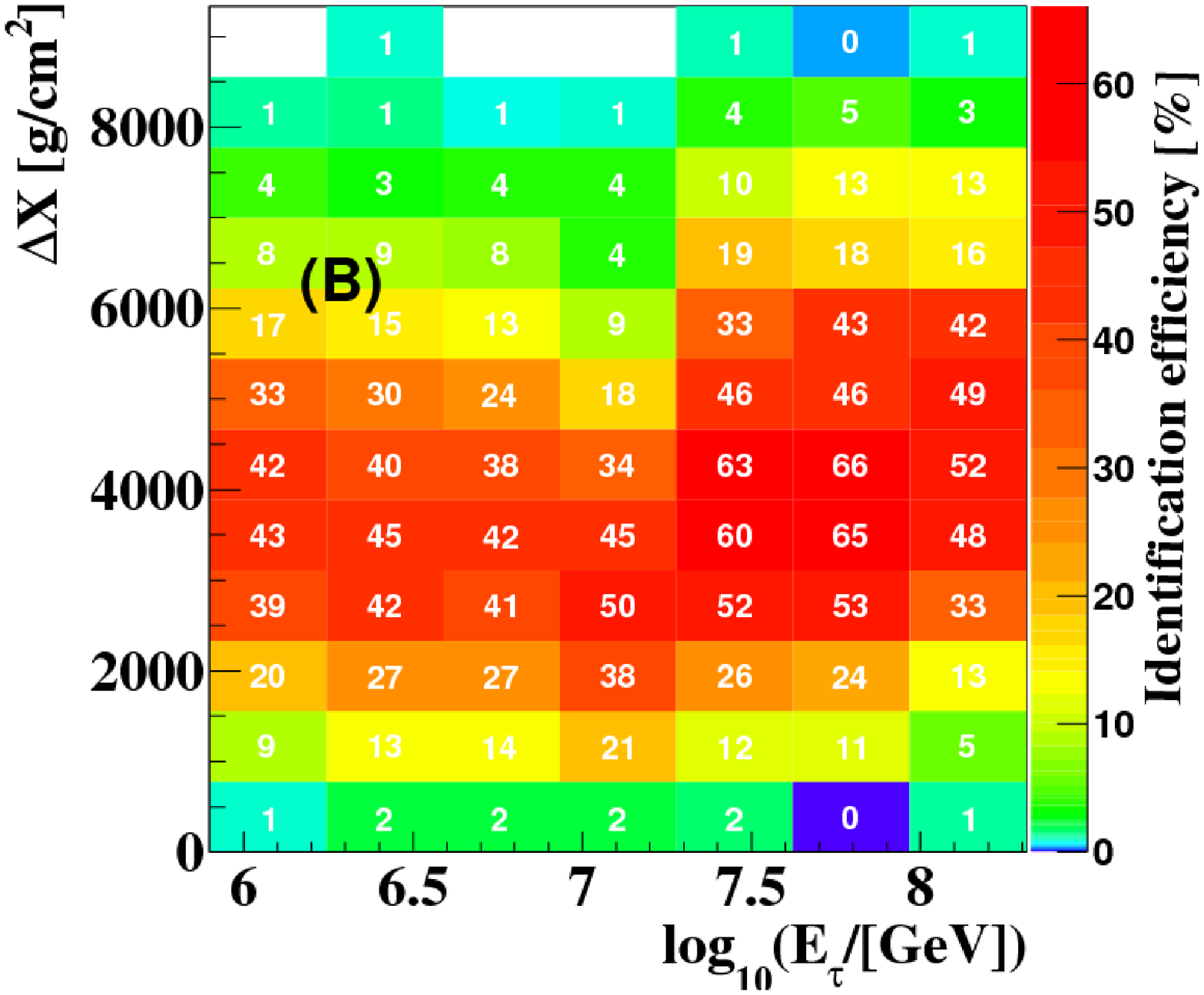}
  \end{center}
 \caption{\label{trigger3333} (A) Trigger and identification efficiency for 1 PeV $\tau$-induced shower  with zenith angle $\theta=87^{\circ}$ for IACT-4 and CTA-E, (B) Identification efficiency for CTA-E as a function of lepton tau energy and injection slant depth measured from detector level.}  
\end{figure*}  

Each  simulated event recorded and calibrated consists of a number of photoelectrons collected
by each pixel in the camera while the trigger gate is opened. The standard trigger configuration requires 
at least three connected pixels with a signal above the discriminator
threshold. However, most of the camera pixels collect light not from the Cherenkov 
shower but from  background. To eliminate the background contribution  an 
 image cleaning is performed~\cite{simtelarray}. The resulting cleaned shower image contains 
only the pixels considered to have physical information on the shower development. 
 The cleaned camera image  is characterized by a set of image parameters introduced by M. Hillas in~\cite{hillas}.
 These parameters provide a geometrical description of the images of showers and are used to infer the energy of the primary particle, its arrival direction and to distinguish between $\gamma-$ray showers and hadronic showers. It is interesting to  study  these parameters also  in the case of  deep $\tau$-induced showers.

In  Figure~\ref{fig::hillas} the distribution of Hillas parameters for deep $\tau$-induced showers  are shown, in comparison to the one  obtained for $p$ and $\gamma-$ induced showers. In general, these parameters  depend on the  geometrical distance of the shower maximum to the detector, which for deep $\tau$-induced showers is much smaller than for inclined  $p$ and $\gamma$-induced showers   which develop  at  the top of the atmosphere. For example, at  $\theta>80^{\circ}$  this distance is  about  a few  hundred kilometers for particles interacting at the top of the atmosphere and only a few tens kilometers for deep  $\tau$-induced showers.  This geometrical effect leads to a rather  good separation of close ($\tau$-induced) and far-away ($p$, $\gamma$) events in the Hillas parameter phase space. This is evident in
the $Size$-parameter, see Figure~\ref{fig::hillas} (A). This parameter  measures  the total amount of detected light (in p.e.) in all camera pixels, and  it  is correlated with the primary energy of the  shower. The size distribution for $\tau$-induced showers is shifted to larger values compared to $\gamma$ and $p-$ induced events, due to closer distances to the detector.

The difference is also seen for parameters characterizing the longitudinal and lateral shower development like {\it Length}  and {\it Width}, Figure~\ref{fig::hillas} (B) and (C). For showers induced by hadron (proton) the  image on the camera is more irregular and is typically  larger compared to  showers induced by photons.  Thus,   the average  value of {\it Length} and {\it Width}  for photons  is expected to be smaller than for protons. At larger inclinations, the so-called the  $\gamma$/hadron separation is weaker since  images become smaller in size. However,  still  the  difference  between $\gamma$ and $p-$induced  showers  is  well visible  in our simulations.\footnote{The peak in the {\it Length} distribution around  $1^{\circ}$ comes from   single muons, which create a ring/arc in the camera and lead  to a large value of length from reconstruction. An example of this   class of events  is shown in  Figure~\ref{fig::hessimage} (Left). }.
 For  $\tau-$induced showers, the maximum of {\it Length} and {\it Width} distribution lies somewhere between the maximum for  $\gamma$ and protons.  
This can be explained by the fact, that the lepton tau decays according to different decay channels~\cite{dpg},   and a $\tau-$induced  shower is usually a superposition of electromagnetic sub-showers  coming from  decays of neutral pions and hadronic  sub-shower  coming from decaying of charged pion. Thus, 
the shower image in the camera can have different topologies i.e. it can look like $p-$events or $\gamma-$events.

The  angular  distance between  the center of the shower image and the camera center  is  called
the   {\it Distance}-parameter. It is correlated with the angle  between the  shower and the telescope axis, and
for larger zenith angles  it  decreases due to larger  detector-to-shower distance. 
However, this parameter  can  also increase when the  detector-to-shower distance  becomes  smaller for fixed zenith angles. 
The effect it well seen in  Figure~\ref{fig::hillas} (D) in  case of point-like simulations  (i.e. when a normal mode of CORSIKA simulations without VIEWCONE option was used),
when the ellipse center of the shower image for deep $\tau$-induced showers compared to $\gamma$-induced  showers moves away from the camera center. 
For the proton simulations, when the direction of primary protons within 5$^\circ$ around 87$^{\circ}$ was varied, the distribution  is shifted to higher  values of distance parameter.   
Three  peaks  seen at {\it Distance} distributions, as for example 
at  2.5$^{\circ}$, 4$^{\circ}$ and 5$^{\circ}$ for $\tau$-induced showers, 
 are due to structures of CTA-E array, which consists of three  different  telescopes types with different FOV.

Another Hillas parameter, which  describes the orientation of the shower image on the camera  according to its center is  {\it Miss}-parameter. 
As we can see in Figure~\ref{fig::hillas} (E)  the distribution  for $\tau$-events   is  shifted to  lower values compared to $p$ events, showing
that this observable has also a strong separation power from the background of hadronic events.

Figure~\ref{fig::hillas}(F) shows  the distribution  of  {\it Alpha}   for  deep $\tau-$,  $\gamma$ and  $p-$induced showers.  {\it Alpha} is the
 angle between the major axis of the ellipse and the direction from the image Center of Gravity  to the center of camera. This parameter has the highest $\gamma$/hadron separation power (for single IACT observation data), since $\gamma$-ray induced images point to the position of the source in the camera, thus they are characterized by a small value of  {\it Alpha}. On the contrary, hadronic showers are distributed isotropically in the sky implying a rather flat {\it Alpha} distribution. However, at large zenith angles    $\gamma$-induced   images have a rather circular shape~\footnote{ For $\gamma$-induced shower at large zenith angle, the Cherenkov  light due a long optical path trigger  only  a few pixels, thus the shape  of the image is less well determined in terms of Hillas parameters. In  order to see  the elliptical structure of  typical $\gamma$-induced  showers
 we need  camera  with a pixel size  much smaller  than what proposed right now for CTA (i.e. between $0.09^{\circ}$  and $0.25^{\circ}$).} rather than an elongated elliptical one implying that 
the {\it Alpha}-parameter is less well determined. At  zenith  angles $\sim87^{\circ}$ the distribution  is quite flat and becomes similar to  the distribution for $p-$events. For  deep $\tau$-induced  showers the distribution  peaks at  small values of {\it Alpha}, showing  a strong separation power from the  
background of hadronic events.

$Distance$, $Miss$ and $Alpha$  only slightly depend on the primary particle energy (as it is shown in Figure~\ref{fig::hillas2} (D-F)  and  the shower zenith angle  (in the range above $80^{\circ}$)). However, as expected, for  energy dependent parameters like: $Size, Lenght, Width$  we observe the expected shift to higher values
 for higher primary particle energies. It is also worth to mention, that  the largest  differences between deep $\tau$-induced showers and  $p$ and $\gamma$-induced  showers are  observed  for the $Size$, $Miss$ and $Alpha$ parameter. Such observables   can be used  to  distinguish deep $\tau$-induced showers from the background of inclined hadronic showers.

In order to evaluate the best set of cuts to identify deep $\tau$-induced neutrino showers, we used the program GARCON~\cite{garcon}, returning the cuts yielding the maximal signal efficiency with minimal background contamination. We considered  a six  parameter phase space 
$\vec{x} = \{Size, Length,Width, Distance, Miss, Alpha\}$. For signal we considered deep $\tau-$ induced showers
(with $\Delta X < 4000$ g/cm$^2$ i.e. $\sim 50$ km from the  detector  and  $\theta=87^{\circ}$). As a source of background we considered 
showers, initiated by primary protons with energies between 1 PeV and 1000 PeV with a differential spectral index of $\gamma=-2.7$,  and interacting at the top of the atmosphere,  with  $\Delta X > 11400$ g/cm$^{-2}$  and  zenith angle $\theta=87^{\circ}$ \footnote{For zenith angles:
$85^{\circ}$, $83^{\circ}$ and $80^{\circ}$, the Hillas distributions looks similar, except the $Size$ distribution for which we observed  a small shift of maximum
to  higher values, when the zenith angles decreases.}.
The  set of optimized cuts retaining most signal and zero left protons  are  listed  in  Table 1 for IACT-4 and CTA-E.
\begin{table*}[ht]
\vspace{-0.0cm}
\small
\center
\begin{tabular}{cccccccccc}
\hline 
array&E$_{i}^{\tau}$ & $Size$ &  $Length$& $Width$&$Distance$ & $Miss$ & $Alpha$ &  Signal Efficiency  \\ 

type&[PeV]  & [p.e.] & [deg] & [deg] & [deg]& [deg] & [deg] & [\%]  \\ 
\hline 
\hline
IACT-4& 1& $>$ 2010  & $<$ 1.81 & $<$ 0.17  &$<$ 0.91 & $<$ 0.15  & $<$ 51  &  31    \\ 
CTA-E && $>$ 791 & $<$  0.35 & $<$ 0.10 &$<$ 2.34 & $<$  0.35 & $<$ 62 & 32 \\     
   & &   &   &   &   &   &  &  &\\  
IACT-4&10 & $>$ 11500 & $<$ 0.52  &  $<$ 0.47 &$<$ 1.09 & $<$ 0.27 & $<$ 90  & 33 \\ 
CTA-E & & $>$ 2590 & $<$  0.39 & $<$ 0.20 & $<$ 3.47  & $<$ 0.66 & $<$ 19 & 27\\ 
& &   &   &   &   &   &  & & \\ 

%
IACT-4 &100 & $>$ 43100  & $<$ 0.71 & $<$ 0.72 &$<$ 2.26 & $<$ 0.131&  $<$ 17& 30 \\ 
CTA-E & & $>$ 8700 & $<$ 0.39 & $<$ 0.30 & $<$ 3.47 & $<$ 0.66 &  $<$ 19 &  27 \\ 
\hline 
\end{tabular} 
\caption{Chosen cuts for the  identification of $\tau$-induced showers and  zenith angle $\theta = 87^{\circ}$. } 
\vspace{-0.5cm}
\end{table*}
 The selection cuts presented in Table 1  (and also in Figure~\ref{fig::hillas}) demonstrate that background events triggering the IACT/CTA telescopes when pointing below (or close to) the horizon can be distinguished from MC neutrino signatures. This criterion  gives a possibility to identify tau neutrinos from the background of hadronic showers  and can be used to calculate the identification efficiency for $\tau$-induced showers.

In Figure~\ref{trigger3333} (A) the  influence of cuts on the trigger probability is shown, while Figure~\ref{trigger3333} (B) 
gives the  identification efficiency  as a function  of the primary energy of the  tau lepton. 
At  vertical depths smaller than $\Delta X < 3000$ g/cm$^2$, we have lower  values of identification efficiency  for  CTA-E  than IACT-4
due  to  the different  altitudes of detectors  i.e.  a higher altitude for CTA-E of 200 m. However,  the  CTA-E distribution is extended to  higher values of distance to the detector, up to $\Delta X=8000$ g/cm$^2$.

\subsection{Event rate calculations }
\label{eventrate}

The total observable rates (number of expected events) were calculated as $N=\Delta T \times \int_{E_{\mathrm{th}}}^{E_{\mathrm{max}}} A^{\mathrm{PS}}(E_{\nu_\tau})\times\Phi(E_{\nu_\tau})\times dE_{\nu_\tau}$, where $\Phi(E_{\nu_\tau})$ is the neutrino flux,  $\Delta T$ the observation time
and $A^{\mathrm{PS}}(E_{\nu_\tau})$  the  point source acceptance. The acceptance for a point source can be estimated as the ratio between the diffuse acceptance  $A(E_{\nu_\tau})$ and the solid angle $\Delta \Omega$ covered by the diffuse analysis, multiplied by the fraction of time the source is visible $f_{\mathrm{vis}}(\delta_{s},\phi_{\mathrm{site}})$ i.e. is given by: $A^{\mathrm{PS}}(E_{\nu_\tau})\simeq A(E_{\nu_\tau}) / \Delta \Omega \times f_{\mathrm{vis}}(\delta_{s},\phi_{\mathrm{site}})$. The fraction of time  where source is visible depends on the source declination ($\delta_{s}$) and the latitude of the observation site ($\phi$).

In this work, the detector  diffuse  acceptance for an initial neutrino energy $E_{\nu_\tau}$ is calculated from:
\begin{eqnarray}
  A(E_{\nu_\tau})  =N_{\mathrm{gen}}^{-1} \times \sum_{i=1}^{N_{k}}
   P_{i}(E_{\nu_\tau},E_{\tau},\theta) \nonumber \\ 
   \times T_{\mathrm{eff},i}(E_{\tau},x,y,h,\theta) \times
   A_i(\theta)\times \Delta \Omega,
\label{aperture}
\end{eqnarray}
where $N_{\mathrm{gen}}$ is the number of generated neutrino events. $N_k$ is the number of $\tau$ leptons with energies $E_{\tau}$ larger than the threshold energy $E_{\mathrm{th}}=1$\,PeV and a decay vertex position inside the interaction volume\footnote{Only tau leptons  which decays in the interaction volume are considered, so the tau decay probability  is included.}. $P(E_{\nu_\tau},E_{\tau},\theta)$ is the probability that a neutrino with energy $E_{\nu_\tau}$ and zenith angle $\theta$  produces a lepton with energy $E_{\tau}$ (this probability was used as "weight" of the event). $A_i(\theta)$ is the physical cross-section of the interaction volume seen by the neutrino. $T_{\mathrm{eff}}(E_{\tau},x,y,h,\theta)$ is the trigger efficiency for tau-lepton induced showers with the decay vertex position at ($x$, $y$) and height $h$ above the ground.

As we already mentioned, in our previous work~\cite{gora:2015}  we assumed an average trigger efficiency of $\langle T_{\mathrm{eff}} \rangle =10$\%  in the  energy range 1-1000 PeV. However, as  seen for example from  Figure~\ref{trigger3333} (A)  the  average  trigger efficiency is significanly
larger than 10\%, even for 1 PeV tau leptons. For  tau  leptons interacting below 4000 g/cm$^2$ with energy in  the range 1-1000 PeV the average trigger efficiency is  about ~90\%(77\%) for IACT-4/(CTA), thus we  also expect a  larger  acceptance  and  event rates by a factor 9 to 8 compared  to what was shown in~\cite{gora:2015}.

\begin{figure}[ht]                                                                                                                
  \begin{center}                                                                                     
    \includegraphics[width=\columnwidth]{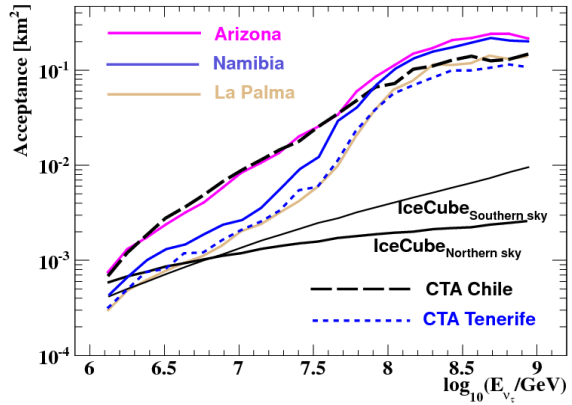}
  \end{center}
\vspace{-0.5cm}
  \caption{\label{fig::acccc} Acceptance for point sources, $A^{\mathrm{PS}}(E_{\nu_\tau})$ to  earth-skimming tau  neutrinos as estimated for the IACT sites and a
 future locations of Cherenkov instrument (Chile)  and  IceCube (as extracted from~\cite{IC-80-acc}). For the Arizona, Namibia and La Palma sites
 the acceptance is calculated with the trigger efficiency  obtained for IACT-4, while for  Chile and Tenerife  the CTA-E trigger efficiency was used instead. The local topographic condition are included. } 
\end{figure}

In Figure~\ref{fig::acccc}  we show our new estimates for the  acceptance  to $\tau$ neutrinos  for  different  IACT-4 sites: La Palma (MAGIC), Namibia (H.E.S.S.) and Arizona (VERITAS)  and  recently chosen  locations of CTA for the North: Chile,  (Armazones: Latitude $\phi=24.58^{\circ}$ S, Longitude $\lambda=70.24^{\circ}$ W)
or Tenerife ($\phi=28.27^{\circ}$ S,  $\lambda=16.53^{\circ}$ W).  As expected,  the acceptance   depends  on local topographic conditions with  the largest
acceptance  for Arizona  and  Chile site ~\footnote{Due to the lack of results from IceCube in the tau-neutrino channel, we use IceCube's muon neutrino acceptance \cite{IC-80-acc} for a sensitivity comparison. This is motivated by the fact that at the Earth we expect an equal flavor flux from cosmic neutrino sources due to full mixing \cite{mixing}. In \cite{up-icecube,diff-icecube} it is also shown that for neutrino energies between 1\,PeV and 1000\,PeV, the muon-neutrino acceptance is only slightly larger than that for tau neutrinos.}.

To calculate the  acceptance for up-going  $\tau$-induced showers we used  the trigger  efficiency instead of the shower identification efficiency, 
since in the studied angular range ($90^{\circ} <\theta < 105^{\circ}$) the expected background from protons and photons  will be negligible. This is also
expected in case of Cherenkov telescopes observations in the direction of mountains, when they are shielded against cosmic rays and star light.
However, in some cases like for example for La Palma  or Tenerife Cherenkov telescopes can be  pointed to the sea.  
Thus,  for high energies ( $>$ 1 PeV) we can expect  a non zero background  component due  to  the presence  of   high energetic  muons or muons bundles
(as for example seen by IceCube~\cite{icecubemuons}) or even gamma showers induced by interacting muons via  bremsstrahlung  or pair production~\cite{kiraly,sciutto}. If the identification efficiency is used instead of the trigger efficiency, the calculated acceptance for IACT-4/CTA  and  the expected event rate  is of about  two/three times  lower.

In Table~\ref{tab::rate222} the expected event rates for IACTs, Tenerife and  Chile site compared to that of IceCube is shown for 
fluxes used in our previous work~\cite{gora:2015}. The rate is calculated for tau neutrinos  with zenith angles  between 90$^{\circ}$ and 105$^\circ$ assuming that the source is in this  FOV  for a period of 3 hours. The Flux-1 and Flux-2 are predictions for neutrino from $\gamma$-ray flare of 3C 279~\cite{2009IJMPD}. Flux-3 and Flux-4 are  predictions for PKS~2155-304 in low-state and high-state, respectively~\cite{Becker2011269}. Flux-5 corresponds to a prediction for 3C~279 calculated in~\cite{PhysRevLett.87.221102},  and it is  at  a similar level in the PeV energy range like  the flux reported by IceCube in case of astrophysical high-energies  neutrinos~\cite{lasticecube}. For   Flux-3 and Flux-4  (i.e. those models covering the energy range beyond   $\sim 1\times10^{8}$ GeV) the event rate  is   a factor  16 to 30 larger what expected for IceCube in the northern sky assuming three hours of observation. For neutrino fluxes covering the energy range below $\sim 5\times10^{7}$\,GeV (Flux-1, Flux-2, Flux-5), the number of expected events for these sites is  at least three times larger  (La Palma) or  seven times larger (Arizona) to what estimated for IceCube. 

\begin{table*}[bt!]
  \caption{\label{tab::rate222} {Expected event rates for Cherenkov detectors  at  different sites compared to  IceCube. The values are calculated with the ALLM~\cite{allm} tau energy loss model and the GRV98lo~\cite{GRVlo} cross-section, with $f_{\mathrm{vis}}=100$\%, $\Delta
 \Omega=2\pi (\cos(90^{\circ})-\cos(105^{\circ}))=1.62$ and $\Delta T=3$ hours. Rates  are in units $10^{-3}$. For Arizona, Namibia and La Palma site
 the rates are calculated with the trigger efficiency  obtained for IACT-4, while for  Chile and Tenerife with the trigger efficiency obtained for the CTA-E.}}
\begin{center}
\begin{tabular}{cccccccc}
\hline
\hline
 &Flux-1  &Flux-2&  Flux-3 & Flux-4 &Flux-5 \\
\hline
\hline
$N_{\mathrm{La Palma}}$  &2.5 & 1.4 & 0.77 &7.7  &2.3\\
$N_{\mathrm{Namibia}}$   &4.3 & 2.3 & 0.99 &9.9 &3.8 \\
$N_{\mathrm{Arizona}}$   &7.4 & 3.4 & 1.44 &14.4 &6.2 \\
&    &   &      &  & &\\
$N_{\mathrm{Tenerife}}$  &3.0 & 2.2 & 0.73  &7.3 &2.8\\
$N_{\mathrm{Chile}}$  &7.9 & 3.3 & 0.98  &9.8 &6.0\\
&    &   &      &  & &\\
$N^{\mathrm{Northern \mbox{ } Sky}}_{\mathrm{IceCube}}$&  0.68&  0.25 &  0.046  & 0.46 & 0.88 \\
$N^{\mathrm{Southern \mbox{ } Sky}}_{\mathrm{IceCube}}$&  1.1&  0.32 &  0.076  & 0.76 & 0.88 \\

\hline
\hline
\end{tabular}
\end{center}
\end{table*}

The influence on the expected event rate arising from uncertainties on the tau-lepton energy  loss and  different neutrino-nucleon cross-sections was studied in our previous work~\cite{gora:2015}. The influence of  systematic uncertainties  on the event rate was estimated to be  about  +14\%/-7\% for Flux-1 and +43\%/-16\% for Flux-3.

\section{Summary}
\label{summary}
 In this paper,  we present results of MC simulations  of   $\tau$-induced air showers  for IACTs  and  for selected  CTA  arrays.  We  calculated the  trigger and identification efficiencies  for $\tau$-induced showers and  study the  properties of their  images  on the camera  focal plane,  as described by Hillas parameters.   In  our previous work~\cite{gora:2015}, which assumed a trigger efficiency of 10\% we predicted, that  the  calculated  neutrino  rates  are  comparable or even larger (above $\sim30$ PeV)  to  what  expected  for  the  IceCube neutrino telescope assuming  observation times for  Cherenkov  telescopes  of  a  few  hours. In this work we have carried out  more realistic simulations  and we predict even larger efficiencies expected for IACTs. In the most favorable case in Table 2, 
we expect  1 event during  210 hours of observation. Taking into account that for this purpose IACTs have to be pointed below the horizon during moonless nights, the detection of tau neutrinos seems to  be difficult. However, such  observation time/or even larger  can be an accumulated  during periods with high clouds,  when those instruments are normally not operated. Very often  (for example for the La Palma  site this is of about 100 hours/year) high clouds prevent the observation of $\gamma$-ray sources, but still allow  pointing the telescopes to the horizon. 
This makes the perspective of detection tau neutrino induced shower by IACT more attractive.

\end{document}